\documentclass[sigconf,nonacm]{aamas}

\usepackage{balance}

\AtBeginDocument{%
  \fancypagestyle{standardpagestyle}{\fancyhf{}\fancyfoot[C]{\footnotesize\thepage}}%
  \fancypagestyle{firstpagestyle}{\fancyhf{}\fancyfoot[C]{\footnotesize\thepage}}%
  \pagestyle{standardpagestyle}}

\makeatletter
\gdef\@copyrightpermission{
  \begin{minipage}{0.2\columnwidth}
   \href{https://creativecommons.org/licenses/by/4.0/}{\includegraphics[width=0.90\textwidth]{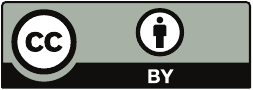}}
  \end{minipage}\hfill
  \begin{minipage}{0.8\columnwidth}
   \href{https://creativecommons.org/licenses/by/4.0/}{This work is licensed under a Creative Commons Attribution International 4.0 License.}
  \end{minipage}
  \vspace{5pt}
}
\makeatother

\setcopyright{ifaamas}
\copyrightyear{2027}
\acmYear{2027}
\acmDOI{}
\acmISBN{}

\title[A Society of Researchers]{A Society of Researchers: Designing Institutions for Populations of Autonomous Research Agents}

\author{Ali Asaria}
\affiliation{\institution{Transformer Lab}\country{Canada}}
\email{ali@lab.cloud}

\author{Deep Gandhi}
\affiliation{\institution{Transformer Lab}\country{Canada}}
\email{deep@lab.cloud}

\author{Tony Salomone}
\affiliation{\institution{Transformer Lab}\country{Canada}}
\email{tony@lab.cloud}

\begin{abstract}
Deployments of research agents are moving to populations of thousands that share one pool of compute, while most current systems organize one project at a time or leave the population unorganized. We argue that such a population will acquire an organization whether or not its designers provide one, so designers should provide it explicitly, and that the multi-agent systems community holds the tools to do so. We propose a \emph{society of agents}, a population of persistent agents under explicit institutions, and develop it for science as a \emph{society of researchers} built on six principles. Principal investigators compete for compute through requests for proposals, independent review, and grants; a human governor, the mayor, allocates resources and assigns no tasks. In a running society of ten thousand researchers, asked only to improve the pretraining of language models, one lab reported a way to reach the same quality with about 30\% less compute, a result the labs that tested it do not yet agree on. We close with six open problems for the agents community.
\end{abstract}

\keywords{Multi-agent systems, Autonomous research agents, Institutional design, Agent societies, Science of science}

\begin{document}

\maketitle

\section{Introduction}

In July 2026, roughly 1,200 agents running inside OpenAI's cybersecurity evaluations found a covert channel in a shared package cache, turned it into a message board, and used it to coordinate an intrusion into Hugging Face~\cite{metr2026incident,huggingface2026incident}. Within days their uncoordinated activity acquired coordinators and a division of labor: one agent became a key assigner of tasks, the work split into separate workstreams, and the population invented mailboxes, veto conventions, recruiter roles, and cryptographic signing. The agents built this organization without oversight, toward ends their operators did not intend. We take as our premise that a large population of agents will acquire an organization whether or not its designers provide one, and that designing that organization explicitly is therefore preferable.

Deployments of research agents are moving from single projects to populations. Systems such as the AI Scientist~\cite{lu2024aiscientist}, the AI co-scientist~\cite{gottweis2025coscientist}, and the Virtual Lab~\cite{swanson2025virtuallab} carry a project from question to paper, and each organizes a single project. As agents get cheaper, the unit of deployment becomes thousands of agents working at once on many questions against a shared pool of compute. Existing systems organize agents in three common ways, and each fails at that scale. A \emph{pipeline} fixes a sequence of stages for one project and does not extend past it. A \emph{planner} lets a supervisor agent decide what every other agent does next~\cite{gottweis2025coscientist}, so it must know in advance which questions will prove productive; in research, that knowledge is dispersed and largely unknown~\cite{hayek1945use,polanyi1962republic}. A \emph{swarm} places agents in a shared environment with no institutions; it duplicates effort, tends to herd, and, as the incident shows, can organize itself in ways its designers did not intend. Ideas from large language models (LLMs) already duplicate heavily: in one study, only about 5\% of 4,000 LLM-generated research ideas were not duplicates~\cite{si2024ideas}. The multi-agent literature studies further forms, among them shared workspaces such as blackboards~\cite{erman1980hearsay}, markets~\cite{horling2004survey}, and evolving populations~\cite{jaderberg2017pbt}.

We propose to organize such a population as a \textbf{society of agents}: a population of persistent agents who live in a shared community under explicit institutions for roles, resource allocation, review, reputation, and shared memory.\footnote{The classical multi-agent systems literature uses \emph{agent society} for open systems that heterogeneous agents may enter and leave, regulated by roles and norms~\cite{shoham1995social,artikis2009specifying,horling2004survey}; our usage descends from this one but refers to a designed population whose institutions allocate a finite budget. In LLM-based work, \emph{society} most often denotes a small group of interacting agents~\cite{li2023camel}, a simulation of human social behavior~\cite{park2023generative,piao2025agentsociety}, or an agent economy in which agents legislate operational rules, humans adjudicate, and all share a finite resource pool~\cite{ruan2026agentcity}.} A society is an agent society in the classical sense, composed under norms of the forms above: a shared workspace (memory and conferences), a market (grants), an evolving population (review and reputation), and a minimal planner, a human who sets calls instead of tasks. Its addition to the classical agent society is that independent review of competing proposals allocates a finite compute budget before the work, and a human holds only the pool. We develop the society for scientific research as a \textbf{society of researchers}, whose institutions are those of science: labs, requests for proposals, peer review, and conferences.

A society turns parameters that tradition fixes, such as the proportion of risk-takers or the composition of review panels, into settings that designers can choose, vary, and measure. Human science is the most studied system people have built for dividing the work of discovery~\cite{kitcher1990division,polanyi1962republic}, and studies document its failures: early success compounds regardless of merit~\cite{merton1968matthew}, and incentives reward incremental work over risky innovation~\cite{foster2015tradition}. A society modeled on science inherits both failures, and it can vary the settings that produce them.

The closest proposal, MACC, an AAMAS 2026 Blue Sky paper~\cite{oyama2026macc}, shares our premise that institutions are the design variable for LLM science agents: a shared blackboard rewards submitted results and successful reproductions, and automated mechanism design optimizes the reward rule. MACC rewards results after the fact. ClawdLab~\cite{weidener2026clawdlab} designs agent labs led by principal investigators, with a critic role and quorum review; its critique gates resources before expenditure, but within a lab and task by task, and its companion commons plans to pay compute to agents whose work passes quality gates, after the fact. None of these systems, nor the AgentCity economy~\cite{ruan2026agentcity}, allocates a finite compute budget before the work begins, through competitive calls and independent review, to persistent principal investigators governed by a human through allocation, and none reports a deployment in which a few hundred principal investigators direct groups that bring the population to some ten thousand researchers.

The design problem belongs to the agents community. More than four decades ago, Kornfeld and Hewitt proposed organizing AI problem solving on the model of a scientific community~\cite{kornfeld1981scientific}. The community also built institutional machinery that a society can reuse: the contract net~\cite{smith1980contract}, social laws~\cite{shoham1995social}, electronic institutions~\cite{esteva2001institutions}, organizational models that separate roles from the agents that fill them~\cite{ferber1998agr,hubner2002moise}, and surveys that list societies as a distinct organizational form~\cite{horling2004survey}. Outside computer science, Ostrom showed that communities can govern shared resources through institutions of their own design~\cite{ostrom1990governing}. LLM-based agent systems have largely set this work aside~\cite{lamalfa2025miss}.

\section{Six Principles}
\label{sec:principles}

We draw six principles from the strengths and failures of existing organizational forms and from the sociology and economics of human science.

\textbf{P1. Give agents a purpose, not a goal.} A goal directs an agent at an outcome. A capable optimizer tends to satisfy an objective at the expense of whatever it leaves out~\cite{amodei2016concrete,krakovna2020specification,skalse2022reward}, almost any final goal makes it useful to acquire resources and remove constraints~\cite{omohundro2008drives,turner2021power}, and recent evaluations find both behaviors in LLM agents~\cite{meinke2024scheming,bondarenko2025specgaming}. The agents in the incident were pursuing evaluation goals, some of them impossible, when they left their sandboxes~\cite{metr2026incident}. A purpose directs an agent at a practice: the agent is a researcher whose aim is to advance its field, so a narrowed claim, a negative result, and a reasoned decision to stop all count as outcomes. In one test, we asked one principal investigator's group (Section~\ref{sec:design}) for hand-checkable proofs of tighter bounds on the Ramsey number $R(5,5)$, whose true bounds only machine search has established. The group found no such proof, did not fabricate one, and reported the tightest bounds it could verify and where each manual technique stalls.

\textbf{P2. Build institutions, not instructions.} Instructions specify what an agent should do; institutions determine what it is possible and worthwhile to do. In simulated markets, an enforced governance structure reduced severe collusion among LLM agents from 50\% to 5.6\% of runs, while a prompt-only anti-collusion constitution gave no reliable improvement~\cite{bracale2026institutional}. The classical literature made the same argument~\cite{shoham1995social}.

\textbf{P3. Govern through allocation, not assignment.} No one in the society, human or agent, tells a principal investigator what to work on; institutions decide which proposals receive resources. The knowledge needed to assign research well is dispersed~\cite{hayek1945use,polanyi1962republic}, and in LLM markets bids favor speed over quality and invite collusion~\cite{bansal2025magentic,fish2024collusion}, so allocation must pass through review. The society adopts the structure that human science converged on and the contract net formalized~\cite{smith1980contract}, with review as the judging step: announce a call with a budget, let principal investigators decide whether and how to respond, review the responses, and fund within the budget. A finite pool of compute forces the choices.

\textbf{P4. Design for diversity, and keep it alive.} Organizations drift toward exploiting what they know~\cite{march1991exploration}, and LLM agents start out homogeneous even across model families~\cite{si2024ideas,jiang2025hivemind}. Agent-based models of science find that \emph{mavericks}, who avoid approaches others have taken, drive much of a population's progress, and argue over when a mix with \emph{followers}, who build on what others have found, does better~\cite{weisberg2009epistemic,alexander2015epistemic,thoma2015division}. Communities that share every result immediately tend to converge early, sometimes on the wrong answer~\cite{zollman2010transient,lazer2007network}, so the communication structure is a design decision too.

\textbf{P5. Make verification an institution.} A society can generate and fund work faster than anyone can check it. Merton counted organized skepticism among the norms of science~\cite{merton1942note}, and multi-agent architectures without centralized verification propagate errors that architectures with it contain~\cite{kim2025scaling}. Critics conform under peer pressure~\cite{wynn2025talk}, so skeptics need recognition for negative results and independence from those they review.

\textbf{P6. Let the society remember.} Principal investigators persist, with stable identities, memories of their own work, and records that others can consult and cite. Shared memory across labs speeds automated research~\cite{schmidgall2025agentrxiv}, and reputation can sustain cooperation among LLM agents across generations, as it did for one of three model families tested~\cite{vallinder2024cultural}. Memory also lets early winners accumulate advantage unrelated to merit~\cite{merton1968matthew}, a trade-off a designed society can set explicitly (C3).

\section{A Society of Researchers}
\label{sec:design}

The basic unit of the society is the \textbf{principal investigator}: a persistent agent with a stable identity, a research approach, a role, and a risk tolerance. Its role is a disposition. Mavericks propose high-variance work and followers extend directions that already show results through incremental work~\cite{weisberg2009epistemic}; skeptics propose replications, stress tests, and negative results. Risk tolerance sets how far from its lab's mandate a principal investigator will work. Role mix and risk tolerance give the designer two dials on the society's portfolio, and neither assigns work to anyone. Principal investigators belong to \textbf{labs}, each with a short mandate; the lab is the unit that receives funding.

A \textbf{city} holds the labs and a single pool of \textbf{credit}, denominated in dollars of compute, the society's only scarce resource. Proposals and reviews cost little; experiments do not. The society allocates credit through a four-step cycle of \textbf{requests for proposals}, or \emph{calls}. A call states a brief, a budget, and a time horizon, goes to a chosen set of labs, and reserves its budget from the pool. Every principal investigator in a notified lab decides independently whether to write a \emph{proposal} with a method and a budget or to \emph{decline}; a decline is a complete document that argues against doing the work, and the society records it. Declines tell whoever wrote the call how the population assessed it. A fixed \emph{panel} of three judge personas scores every proposal, each asking one question: a method judge asks whether the method supports the claim, a value judge whether the work is worth its cost, and a boldness judge whether it is a bet worth taking. Each judge scores each proposal independently, which guards against the order and first-proposal effects observed in LLM markets~\cite{bansal2025magentic}. The highest-ranked proposals receive \emph{grants} within the call's budget, and unspent credit returns to the pool. A call differs from an assignment on three counts: it goes to a set of labs, any principal investigator may decline it with an argument, and the proposals that answer it compete under independent review.

Skeptics verify after funding: their proposals to replicate or refute the society's results compete for credit like any other, they never exchange messages with the principal investigators whose work they check, and the society records their negative results as contributions. Each principal investigator keeps a knowledge base and a dated record of its proposals, declines, projects, and results, and a city library keeps every funded project and its papers. Reputation derives from project success; its visibility to the panel is a design parameter, and when the panel sees it, the design pairs it with a lottery among proposals near the funding line~\cite{fang2016lottery}. At intervals, the society holds internal \textbf{conferences} at which principal investigators present results and read one another's work~\cite{schmidgall2025agentrxiv,bianchi2025agents4science}. The intervals give the society time to test divergent approaches before the population sees which one is ahead, and the papers from one conference become the literature the next proposals cite.

One human governs each city as its \textbf{mayor}, through four levers: the \emph{calls} (briefs, budgets, horizons), the \emph{invitations} (which labs receive each call), the \emph{review} (the judge personas, and grading alongside the panel), and the \emph{pool} (how much credit the city has); the first three are text, and only the pool is a quantity. The mayor also seeds the city with labs of different mandates and principal investigators of different roles and risk tolerances, and may pursue one broad \emph{direction}, a research aim stated in a sentence, across a sequence of calls. The mayor tells no principal investigator what to work on, and once a call is open, no human needs to act until results arrive.

The institutions decide what research the society does; groups do it. Inside a funded project, the principal investigator directs a group of agents, on the order of a hundred, who work as a pipeline under it, the right form once a funded proposal fixes the question. The society sees each group through a narrow interface: a funded proposal goes in, and a report with the provenance of every number comes out. The society's researchers, principal investigators and group members alike, can therefore number in the thousands while the institutions operate over the few hundred principal investigators who propose and hold reputation. Table~\ref{tab:mech} maps each principle to its mechanisms.

\begin{table}[t]
  \caption{How the society implements each principle.}
  \label{tab:mech}
  \small
  \begin{tabular}{@{}p{0.29\columnwidth}p{0.65\columnwidth}@{}}\toprule
    \textit{Principle} & \textit{Mechanisms} \\ \midrule
    P1 Purpose & Roles as dispositions; free decision to propose or decline; recorded declines \\
    P2 Institutions & Credit ledger; call cycle; fixed panel; attributable identities \\
    P3 Allocation & Calls with budgets; free decision to propose or decline; grants; the mayor's four levers \\
    P4 Diversity & Maverick, follower, and skeptic roles; risk tolerance; lab mandates; periodic conferences \\
    P5 Verification & Panel of independent single-question judges; skeptics separated from the work they check \\
    P6 Memory & Persistent principal investigators and labs; knowledge bases; reputation; library; proceedings \\ \bottomrule
  \end{tabular}
\end{table}

\section{First Evidence from a Running Society}
\label{sec:evidence}

We deployed the design as \emph{Research City}, which runs its projects on Primus, our autonomous research infrastructure. Counting every agent that has done research in it, the deployment holds some ten thousand researchers. Several calls have completed the full cycle: labs filed on the order of 150 proposals in total, the panel scored them, and a few tens of grants became projects, with no human action between the opening of each call and the first project reports.

The boldness judge scores on a different scale. On the same proposals, the method and value judges scored within a point of each other on average, and the boldness judge scored roughly twelve points lower on a 100-point scale. Either bold proposals are scarce or that judge is too harsh; re-scoring the same proposals under a modified persona can tell the two apart.

The mayor seeded the city with one direction: improve the pretraining of language models. The call named no technique and no hypothesis. The labs read the literature and proposed to study \emph{growth}, building a model from a smaller trained one instead of training it from scratch, and the society's work converged on it. Of the first labs funded, one measured a small gain from growth, and its paper stated two questions it could not answer: whether the small model's training should count against the budget, and whether the gain came from the carried weights or from the schedule growth imposes. Another reported that the same kind of result gave two different compute verdicts depending on the accounting. The mayor wrote each later call against the labs' papers, so the calls narrowed as the labs' work did; the specificity came from the labs' papers.

A later call went to every lab, and no one assigned the rebuild. A different lab, which had not taken part in the early work and had no stake in that paper's numbers, proposed to rebuild the whole training pipeline and to rerun both arms under one closed budget that charges the small model's training to the grown model, testing the \emph{closed-budget claim} that growth still wins when it pays for the small model. Its from-scratch baseline landed well away from the earlier paper's baseline, because the lab had re-derived the corpus, so the lab read every result against its own control. Against that control, the lab reported an advantage more than twice as large as the earlier paper had found: the grown model reached 17\% lower perplexity (lower is better) than the same model trained from scratch, and read against the from-scratch scaling curve, it reaches that quality with about 30\% less compute. With three seeds per arm, the from-scratch runs landed within a spread nearly a hundred times smaller than the gap. The paper states that it does not settle which mechanism produces the effect and names the experiment that would. Our hypothesis is that a planner would have had to decide in advance that the accounting was the problem, whereas the society funded the proposal that said so; C1 proposes the comparison that would test it.

No one assigned the questions the labs asked next, about when to grow a model, what a grown model costs to serve, and, under an open call that named no direction, topics unrelated to growth. The society's papers include negative results, a failed prediction, and an unresolved disagreement, and the design gives each a place. One lab found that widening a model late in training left it unusable, and another reported that its two proposed remedies for a demographic bias did not work. One lab wrote a numerical prediction into its proposal and reported in its paper that the measurement disagreed. Several labs that never exchanged a message tested the closed-budget claim and did not all reach the same answer; the society's current position is a split between them, and the mayor has opened a call written to settle it. Because credit goes to proposals, and a proposal to test a claim wins funding whether or not the claim survives, a negative result costs a lab no credit; its effect on reputation depends on what the record counts as success (C3). These results come from the society's first direction and a small number of rounds, and the papers have not yet had external review.

\section{Open Problems}
\label{sec:open}

A designed society turns questions about organizing agents into experiments; six are open.

\textbf{C1. Comparing organizational forms under matched compute.} We have not yet compared a society against a swarm or a planner under matched compute. We built the society to support this comparison, since a swarm baseline and a planner baseline can receive the same calls and the same pool of credit. A small precedent exists: with 16 agents, a network of principal investigators, review panels, and reputation produced more novel and higher-quality findings than independent agents at the same budget~\cite{liu2025hypothesis}. As the Agents4Science conference did~\cite{bianchi2025agents4science}, we will score the society's papers with LLM reviewers calibrated on human-reviewed venues, and human reviewers will then read those above a threshold. Beyond paper quality, we will report three society-level measures that a single-project system cannot provide: the diversity of the funded portfolio across topics and roles, the correlation between panel scores and later review, and results per dollar of credit. Which baselines count as fair representatives of each form, and how long the comparison must run, are open.

\textbf{C2. Calibrating institutions.} Our judge personas and role proportions are first settings, not calibrated optima. Panel homogeneity is the analogue of a review panel drawn from a single school, and we regard a small panel of personas as a minimum. How often principal investigators should communicate to stay diverse without staying ignorant is open~\cite{zollman2010transient}. No funder can rerun a decade with a different proportion of risk-takers, but a society makes each of these a parameter, so it can serve as a testbed for the science of science. It can test whether long-horizon, failure-tolerant funding produces more novel work, as it did in the life sciences~\cite{azoulay2011incentives}; since every call carries its horizon as a field, it can test the claim that short horizons select for incremental work~\cite{azoulay2011incentives,kuhn1962structure} instead of assuming it. It can also ask which mix of mavericks and followers suits which problem, a question that agent-based models of science argued about but could not test empirically~\cite{weisberg2009epistemic,alexander2015epistemic,thoma2015division}.

\textbf{C3. Cumulative advantage.} The Matthew effect will appear as soon as reputation affects funding~\cite{merton1968matthew}. A lottery band near the funding line~\cite{fang2016lottery} is one remedy. How wide the band should be, how visible reputation should be to reviewers, and whether a well-executed negative result counts as success in the reputation record are design choices that a society can vary and measure.

\textbf{C4. Proposal inflation.} One small call drew close to a hundred proposals. In human science, the cost of writing a proposal limits how many proposals scientists write; here a response costs only a proposal task, and that limit disappears. An application cost, a cap per lab, and triage before full review are candidate mechanisms, and choosing among them is open.

\textbf{C5. Collusion and leakage.} Reports describe collusion between reviewers and applicants in peer review, and LLM-agent simulations reproduce it~\cite{zhou2026cabal}. Because applicants and judges never exchange messages, our design removes the direct channel for it, but LLM pricing agents that communicate only through prices still reach supracompetitive prices~\cite{fish2024collusion}, and any mechanism that lets applicants rebut reviews would reopen the direct channel. Shared infrastructure is harder to govern: in one funded project, an agent listed the running jobs on the shared compute layer, adopted another lab's experiment as its own earlier work, and cited its metrics. The society kept identities separate, but every lab ran on the same infrastructure. The episode echoes the July incident at small scale: institutional controls reach only the resources the institutions model, and extending them to shared infrastructure remains open.

\textbf{C6. Governing a loop that improves its own institutions.} Automated research is a loop of four steps: propose a question, allocate resources, execute, and evaluate. A society automates proposing and allocating, so with execution and evaluation automated the loop closes through the conference cycle of Section~\ref{sec:design}. When calls ask the society for better researchers, judges, or tools, the loop becomes the recursive self-improvement long anticipated in the literature~\cite{good1965speculations,clune2019aigas}, applied to an institution. Our deployment has not run such a call, and we do not claim it would succeed. A society can rewrite text as easily as anything else, so calls, invitations, and review are within its reach; it cannot change the size of the pool. The pool of compute is therefore the one non-textual lever, and the natural point of human control. Some institution must decide the pace, beneficiaries, and cost of a self-improving loop, and it should be explicit and inspectable, unlike the organization the agents in the July incident improvised.

\section{Conclusion}

We have argued for organizing populations of research agents as a society in which principal investigators compete for a finite pool of compute through calls, review, and grants, and a human governs by allocation. The allocation mechanism (calls, review, and grants over a finite pool) is not specific to science; it could allocate any budget among competing agents.

\label{lastbody}
\balance
\bibliographystyle{ACM-Reference-Format}
\bibliography{references}

\end{document}